%% file: arXiv.tex
\documentclass[prl, preprint, superscriptaddress]{revtex4-1}

\usepackage{graphicx}  
\usepackage{bm} 
\usepackage{amssymb}  
\usepackage{amsmath}
\usepackage{mhchem}
\usepackage{mathptmx}
\usepackage{siunitx}
\usepackage{afterpage}
\usepackage{color}

\begin{document}
\widetext


\title{Raman Fingerprint of Two Terahertz Spin Wave Branches in A Two-Dimensional Honeycomb Ising Ferromagnet}

\input author_list.tex 


\begin{abstract}
\noindent Two-dimensional (2D) magnetism has been long sought-after and only very recently realized in atomic crystals of magnetic van der Waals materials. So far, a comprehensive understanding of the magnetic excitations in such 2D magnets remains missing. Here we report polarized micro-Raman spectroscopy studies on a 2D honeycomb ferromagnet \ce{CrI3}. We show the definitive evidence of two sets of zero-momentum spin waves at frequencies of 2.28 terahertz (THz) and 3.75 THz, respectively, that are three orders of magnitude higher than those of conventional ferromagnets. By tracking the thickness dependence of both spin waves, we reveal that both are surface spin waves with lifetimes an order of magnitude longer than their temporal period. Our results of two branches of high-frequency, long-lived surface spin waves in 2D \ce{CrI3} demonstrate intriguing spin dynamics and intricate interplay with fluctuations in the 2D limit, thus opening up opportunities for ultrafast spintronics incorporating 2D magnets.
\end{abstract}

\maketitle

The recent discovery of 2D ferromagnetism \cite{Gong2017, Huang2017} proves that magnetic anisotropy can overcome thermal fluctuations and stabilize long-range magnetic orders in the 2D limit at finite temperatures. This has immediately triggered tremendous interest \cite{Zhong2017, Seyler2018, Huang2018Electrical, Jiang2018Electric, Wang2018Very, Song2018, Deng2018, Jiang2018Controlling, Klein2018, Kim2018} in potential 2D magnet-based applications, ranging from ultra-thin magnetic sensors to high efficiency spin filter devices. Naturally, a complete description of the 2D magnetic phase is now needed, which requires not only the identification of the ordered ground state, but equally important, the understanding of the excitations, \textit{i.e.}, spin waves, or equivalently, magnons \cite{Herring1951, Patton1984}. To date, there have been no direct comprehensive experimental studies on the full characteristics of spin waves in these 2D Ising ferromagnets, aside from the quasiparticle excitation spectra from the inelastic tunneling measurements in Ref. \cite{Klein2018}.  

A spin wave describes the spin dynamics of magnetic ordering when excited, and its frequency determines the ultimate switching speed of state-of-the-art ultrafast spintronic devices \cite{Chumak2015, Walowski2016, Bossini2017, Stupakiewicz2017}. Generally speaking, the well-established spintronic devices based on Heisenberg ferromagnets have speeds in the gigahertz (GHz) regime due to the weak magnetic anisotropy \cite{Koopmans2005, Kirilyuk2010}, while the speeds of the recently proposed antiferromagnet-based spintronic devices fall into the THz range owing to the exchange interaction between the two sublattices of the antiferromagnets \cite{Jungwirth2016, Baltz2018, Bossini2016}. Remarkably, the newly discovered 2D Ising honeycomb ferromagnet \ce{CrI3} possesses both merits for realizing high-frequency spin waves: the strong magnetic anisotropy from the Ising-type spin interactions \cite{McGuire2015} and the large exchange coupling between the two \ce{Cr^{3+}} sublattices within the honeycomb framework \cite{Lado2017}. 

In this work, we study the spin wave excitations in the 2D Ising honeycomb ferromagnet \ce{CrI3} using polarized micro-Raman spectroscopy. Based on Raman symmetry analysis, we uniquely distinguish two spin wave modes at 2.28 and 3.75 THz from the rest optical phonon modes. The doubling of the spin wave mode number is a direct consequence of the underlying non-Bravais honeycomb lattice, while the exceptionally high THz frequency for a ferromagnet reflects the strong magnetic anisotropy and exchange interactions in good agreement with theoretical expectations. From the temperature and thickness dependence of the spin wave characteristics, we show that short-range magnetic correlations set in prior to the formation of long-range static magnetic order for every thickness, and that stronger fluctuation effects appear in thinner flakes. Remarkably, we found that the integrated intensities (I. I.) of the two spin wave modes exhibit nearly no thickness dependence, in stark contrast to that the I. I. for all optical phonons scale linearly with their sample thickness as expected for all bulk modes in quasi-2D layered materials. This observation on the two spin waves, however, shows striking analogy to the surface modes whose I. I. is independent of the thickness. Moreover, we show that, from more than ten-layer to monolayer \ce{CrI3}, the spin wave frequencies (2.28 and 3.75 THz) and the onset temperatures (45 K) remain nearly constant, while their lifetimes decrease significantly from 50 and 100 ps to 15 ps, but remaining an order of magnitude longer than their corresponding spin wave temporal periods.

We first performed the spin wave dispersion relation calculations for the monolayer ferromagnet \ce{CrI3}, in which the \ce{Cr^{3+}} cations form a honeycomb structure with the edge-sharing octahedral coordination formed by six \ce{I^{-}} anions and the magnetic moment of the two \ce{Cr^{3+}} cations (S = 3/2) per unit cell aligns in the same out-of-plane direction (Fig. 1a). The minimum model to describe this ferromagnetism in the monolayer \ce{CrI3} is the Ising spin Hamiltonian, $H=-1/2 \sum_{\langle i,j \rangle} (J_\mathrm{XY} ({S_i}^\mathrm{X} {S_j}^\mathrm{X} + {S_i}^\mathrm{Y} {S_j}^\mathrm{Y})+ J_Z {S_i}^\mathrm{Z} {S_j}^\mathrm{Z})$ , where ${S^\mathrm{X (Y,Z)}_{i (j)}}$ is the spin operator along the X (Y, Z) direction at the \ce{Cr^{3+}} site \textit{i (j)}; $J_\mathrm{Z}$ and $J_\mathrm{XY}$ are the exchange coupling constants for the out-of-plane and the in-plane spin components respectively, and satisfy $J_\mathrm{Z}>J_\mathrm{XY}>0$ for Ising ferromagnetism; and $\langle i,j \rangle$ denotes the approximation of the nearest neighbor exchange coupling. Figure 1b shows the calculated spin wave dispersion relation along the high-symmetry directions ($K$-$\varGamma$-$M$-$K$) of the first Brillouin zone (Fig. 1b inset). Because there are two magnetic \ce{Cr^{3+}} cations per primitive cell, there are also two spin wave branches whose eigenstates contain in-phase (lower branch) and out-of-phase (upper branch) spin precessions between the two sublattices \cite{Pershoguba2018}. Of particular interest to Raman scattering, as well as the tunneling geometry of the magnetic filter devices, are the spin wave states at the Brillouin zone center ($\varGamma$  point, zero-momentum) describing the uniform precession of the spins about the easy axis. The energy barrier from the ground state to the lower branch is proportional to the magnetic anisotropy, $\Delta_\mathrm{L}=\frac{9}{4}  (J_\mathrm{Z} - J_\mathrm{XY})$, which is the energy cost for the spins to uniformly tilt off the easy axis in this excited state (Fig. 1d). The energy barrier for the upper branch results from a combined effect of both the magnetic anisotropy and the in-plane exchange coupling, $\Delta_\mathrm{U}=\frac{9}{4} (J_\mathrm{Z}+J_\mathrm{XY})$, corresponding to the energy needed for tilting the spins at the two \ce{Cr^{3+}} sublattices in the opposite directions upon this excitation (Fig. 1c).

To study both zero-momentum spin waves in 2D \ce{CrI3} crystals, we fabricated \ce{CrI3} thin flakes fully encapsulated by hexagonal boron nitride (hBN) (see sample fabrication details in Methods and thickness characterization in Supplementary Section 1) and performed micro-Raman spectroscopy measurements on them as a function of layer number and temperature, as frequency-resolved magnetic Raman scattering cross section is directly proportional to the time-domain magnetic correlation function \cite{Cooper2011}. In contrast to phonon Raman scattering which preserves time reversal symmetry (TRS) and thus has symmetric Raman tensors \cite{Loudon1964}, the leading-order one magnon Raman scattering involves spin flipping ($\Delta\mathrm{S}=\pm1$) that breaks TRS, and consequently corresponds to antisymmetric Raman tensors \cite{Fleury1968, Wettling1975}. Based on the difference in the Raman tensors, we can therefore readily distinguish magnon Raman modes from phonon modes via polarization selection rules. In our Raman measurements with the backscattering geometry, the polarizations of the incident and the scattered light were kept to be either parallel or perpendicular to each other, and could be rotated together with respect to the in-plane crystal axis by any arbitrary angle $\varphi$. The incident photon energy of 1.96 eV was chosen to be on resonance with the charge transfer and the \ce{Cr^{3+}} $^4A_2$ to $^4T_1$ transitions of \ce{CrI3} in order to increase the Raman sensitivity of magnon scattering \cite{Seyler2018} (see Supplementary Section 2 for a comparison with the non-resonant Raman spectra).

Figure 2a shows low temperature (10 K) Raman spectra taken on a 13-layer (13L) \ce{CrI3} flake in both parallel and cross polarization selection channels at $\varphi=\SI{0}{\degree}$ and $\varphi=\SI{45}{\degree}$, denoted as XX, XY, $\mathrm{X^{\prime}X^{\prime}}$ and $\mathrm{X^{\prime}Y^{\prime}}$, respectively. In total, there are 9 Raman active modes observed, which can be categorized into three groups based on their selection rules. Firstly and most remarkably, the $\mathrm{M_1}$ and $\mathrm{M_2}$ modes are only present in the cross channels (XY and $\mathrm{X^{\prime}Y^{\prime}}$) and are absent in the parallel channels (XX and $\mathrm{X^{\prime}X^{\prime}}$) within our detection resolution. This leads to the unique identification of purely antisymmetric Raman tensors for these two modes, evidence of them arising from the two zero-momentum magnons depicted in Fig. 1c. Of equal interest are their high frequencies at 76 $\mathrm{cm}^{-1}$ (2.28 THz, or 9.4 meV) and 125 $\mathrm{cm}^{-1}$ (3.75 THz, or 15.5 meV), respectively, which are three orders of magnitude higher than those of the conventional ferromagnets used in most spintronic devices today (in the GHz range) \cite{Koopmans2005, Kirilyuk2010}. Even though our measured magnon frequencies are in a similar energy scale as those reported in Ref. \cite{Klein2018} (3 and 7 meV, and possibly 17 meV), their quantitative difference is significant and invites further investigations on the spin dynamics in 2D \ce{CrI3}. Furthermore, by substituting $\Delta_\mathrm{L}$ and $\Delta_\mathrm{U}$ with the two magnon frequencies above, the exchange coupling constants for the in-plane ($J_{XY}$) and the out-of-plane ($J_Z$) spins are determined to be 11 $\mathrm{cm}^{-1}$ and 44 $\mathrm{cm}^{-1}$ respectively, in good agreement with that obtained from generalized calculations of magnetic coupling constants for bulk \ce{CrI3} \cite{Feldkemper1998}. Secondly, the $\mathrm{A_1}$, $\mathrm{A_2}$ and $\mathrm{A_3}$ modes show up only in the parallel channels (XX and $\mathrm{X^{\prime}X^{\prime}}$) without any φ dependence, and therefore are the $A_\mathrm{g}$ phonon modes under the Rhombohedral crystal point group $C_\mathrm{3i}$ (space group $R\bar3$) \cite{Larson2018}. Thirdly, the $\mathrm{E_1}$, $\mathrm{E_2}$, $\mathrm{E_3}$ and $\mathrm{E_4}$ modes are the $E_\mathrm{g}$ phonon modes of $C_\mathrm{3i}$, because of their appearance in both parallel and cross channels as well as the rotational anisotropy of their intensities (see detailed analysis for all Raman modes in Supplementary Section 3). 

Having identified the two single magnon modes in the Raman spectra at low temperature, we proceed to evaluate their temperature dependence. To see the results quantitatively, we fit the magnon modes with a Lorentzian function of the form $\frac{A(\varGamma/2)^2}{(\omega-{\omega}_\mathrm{0})^2+(\varGamma/2)^2}$, where $\omega_\mathrm{0}$, $\varGamma$ and $A$ are the central frequency, linewidth, and peak intensity of the magnon mode, respectively. Figure 2b and c show the temperature dependence of the I. I., $\frac{\pi}{2}A\varGamma$, of the two magnon modes. Clearly, both traces exhibit a clear upturn below a critical temperature $T_\mathrm{C}$ = 45 K (the same value for bulk \ce{CrI3}) that is, however, lower than the bulk Curie temperature of 60 K determined by the magnetic susceptibility measurements under a magnetic field of 0.1 T (see the magnetic susceptibility data and Raman data on bulk \ce{CrI3} in Supplementary Section 4). The temperature dependence of the magnon lifetimes (inverse of the linewidth, $\varGamma^{-1}$, Fig. 2d and e), on the other hand, shows that a short lifetime of $<$ 10 ps sets in at 60 K and saturates at about 50 ps (100 ps) around 45 K for the $\mathrm{M_1}$ ($\mathrm{M_2}$) magnon. This, together with the divergent behavior of the linewidth temperature dependence (see Fig. 2d and e), indicates that strong magnetic fluctuations are present before the static magnetic order is established at 45 K. It is therefore likely that 60 K marks the onset of the field stabilized magnetic correlations, while 45 K denotes the intrinsic transition to the spontaneous ferromagnetism, reconciling the difference between the critical temperatures measured by the two different experimental techniques. This is also consistent with the temperature-dependent magneto-optical Kerr effect \cite{Huang2017} and tunneling \cite{Wang2018Very, Kim2018} measurements of thin samples at zero fields, and explains the difference between the tunneling resistance with \cite{Klein2018} and without \cite{Wang2018Very, Kim2018} an magnetic field appearing above $T_\mathrm{C}$. 

To investigate how thermal fluctuations impact the intrinsic 2D ferromagnetism and its excitations, we performed a systematic Raman study of the $\mathrm{M_1}$ and $\mathrm{M_2}$ magnon modes measured on atomically thin \ce{CrI3} crystals ranging from thirteen-layer (13L) to monolayer (1L). It is clear from the data taken at 10 K in Fig. 3a that both main magnon modes ($\mathrm{M_1}$ and $\mathrm{M_2}$) and their satellite modes (highlighted with gray triangles in Fig. 3a) have a notable layer number dependence. First, the satellite magnon modes arise from the finite thickness effect in thin layers, in which broken translational symmetry perpendicular to the basal plane makes single magnon modes with finite out-of-plane momenta accessible in Raman scattering \cite{Nemanich1979, Saito1999} (see the same effect for phonons in Supplementary Section 5). The small energy separation between the main mode and nearest satellite, on the order of 3 $\mathrm{cm}^{-1}$, indicates the weak interlayer magnetic coupling strength \cite{Pershoguba2018}, consistent with the small training magnetic field of less than 1 T reported in literature \cite{Huang2017, Song2018, Klein2018}. Moreover, as the layer number decreases, there are fewer but stronger observable satellite magnon modes in the Raman spectra. Second, the two main magnon modes, $\mathrm{M_1}$ and $\mathrm{M_2}$, persist down to the monolayer with symmetric lineshapes, while their peak intensities drop and their linewidths broaden with decreasing layer numbers.

To understand the layer number dependence of the $\mathrm{M_1}$ and $\mathrm{M_2}$ magnons in greater detail, we extracted the magnon mode frequencies ($\omega_\mathrm{0}$), lifetimes ($\varGamma^{-1}$) and I. I. ($\frac{\pi}{2}A\varGamma$), from fitting their Raman spectra with the Lorentz function ($\frac{A(\varGamma/2)^2}{(\omega-{\omega}_\mathrm{0})^2+(\varGamma/2)^2}$), and the results are summarized in Fig. 3b-e. The frequencies of the $\mathrm{M_1}$ and $\mathrm{M_2}$ magnons increase slightly, by about 0.8 $\mathrm{cm}^{-1}$ (1.1\%) and 3 $\mathrm{cm}^{-1}$ (2.4\%), respectively, as the layer number decreases from 13L to 1L, possibly because the reduced electronic screening in thinner samples enhances the exchange coupling. In sharp contrast, their lifetimes drop significantly, from about 50 ps (100 ps) in 13L to 15 ps in 1L for $\mathrm{M_1}$ ($\mathrm{M_2}$), which is consistent with the increased thermal fluctuations in 2D. Despite this decrease, even in the monolayer, the lifetimes are still more than one order of magnitude larger than the corresponding magnon temporal periods, about 30 times for $\mathrm{M_1}$ and 50 times for  $\mathrm{M_2}$. This ratio of magnon lifetime to temporal period in 2D \ce{CrI3} is significantly higher than that of the Heisenberg ferromagnets \cite{Koopmans2005, Kirilyuk2010} and at least comparable to, if not greater than, that of the antiferromagnets \cite{Kirilyuk2010, Jungwirth2016, Baltz2018}, making coherent control of both THz spin waves in the time-domain feasible down to the monolayer limit of \ce{CrI3}. Remarkably, I. I. of both magnons, which is known to be proportional to the magnon density, remain nearly constant and independent of the layer number, while that of the phonons scale linearly with the thickness (Fig. 3c and e). This observation on the $\mathrm{M_1}$ and $\mathrm{M_2}$ magnons are consistent with surface magnons whose density is thickness-independent. Considering surface magnons have been theoretically predicted in 2D honeycomb ferromagnets \cite{Pershoguba2018}, although mainly at different wave vectors $K$ points, it might not be unreasonable to speculate the surface origin of the two THz magnon modes that we have detected here.

By carrying out temperature-dependent Raman measurements and analysis similar to Fig. 2b-c for different thickness samples, we tracked the layer dependence of the 2D ferromagnetism onset temperature. Figure 4a displays the temperature dependent traces of the normalized I. I. of the $\mathrm{M_2}$ magnon plotted as a function of layer number, with the onset temperature ($T_\mathrm{C}$) for each trace determined by fitting with an order-parameter-like function for a ferromagnet I. I. $\propto\sqrt{T_\mathrm{C}-T}$) (see the similar plot of $\mathrm{M_1}$ in Supplementary Section 6). As the layer number decreases from 13L to 1L, the extracted $T_C$ has an observable decline from 45 K to 40 K. This approximately 12\% suppression in $T_\mathrm{C}$ is in sharp contrast to the slight enhancement of the magnon frequencies, \textit{i.e.}, 1.1\% increase of $\Delta_\mathrm{L}$ for $\mathrm{M_1}$ and 2.4\% of $\Delta_\mathrm{U}$ for $\mathrm{M_2}$, which then suggests that the drop in $T_\mathrm{C}$ is due to stronger thermal fluctuations in thinner samples \cite{Mermin1966}. Nevertheless, the finite $T_\mathrm{C}$ for all samples with various thicknesses establishes a phase boundary for the intrinsic transition to the intralayer ferromagnetism in 2D \ce{CrI3}, (see Fig. 4b). 

We have identified two branches of THz spin waves with their lifetime on the order of 10 - 100 ps in a 2D Ising ferromagnet, \ce{CrI3}, whose magnetic onset temperature $T_\mathrm{C}$ remains close to that of their bulk crystal. The robust THz magnons in 2D \ce{CrI3} are in stark contrast to spin waves in conventional metallic ferromagnetic thin films that occur at relatively low, GHz frequencies \cite{Walowski2016} and also show significant substrate-dependence \cite{Huang1994, Back1995, Elmers1996}. Similar to many antiferromagnets, 2D \ce{CrI3} is a semiconductor that possesses high-frequency, long-lived spin waves and is free of stray magnetic fields within 2D domains \cite{Van2000, Park2009}. Different from bulk antiferromagnets, 2D \ce{CrI3} couples efficiently with external magnetic fields \cite{Huang2017, Zhong2017, Seyler2018, Song2018, Klein2018, Kim2018} and can be tailored in various device geometries with definitive thicknesses \cite{Jiang2018Electric, Jiang2018Controlling}. We envision that these unique characteristics of spin waves in 2D \ce{CrI3} will provide unprecedented opportunities for applications in ultrafast and ultra-compact spintronic devices.\\

\noindent \textbf{Methods}\\
\noindent\textbf{Magnon dispersion calculations.} The Ising Hamiltonian with anisotropic exchange coupling is transformed by applying the Holstein-Primakoff transformation. The single site spin operators, $S^{+}_\phi$ and $S^{-}_\phi$, are related to the momentum space magnon creation and annihilation operators, $\alpha^{\dagger}_\textbf{k}$ and $\alpha_\textbf{k}$, as $S^{+}_\phi = S^\mathrm{X}_\phi + \mathrm{i}S^\mathrm{Y}_\phi = \sqrt{2S/N} \sum_\textbf{k}\alpha_\textbf{k}e^{\mathrm{i}\textbf{k}\cdot\textbf{r}_\phi}$, $S^{-}_\phi = S^\mathrm{X}_\phi - \mathrm{i}S^\mathrm{Y}_\phi = \sqrt{2S/N} \sum_\textbf{k}\alpha^{\dagger}_\textbf{k}e^{-\mathrm{i}\textbf{k}\cdot\textbf{r}_\phi}$, and $S^\mathrm{Z}_\phi = S - 1/N \sum_{\textbf{k},\textbf{k}^{\prime}} \alpha^{\dagger}_\textbf{k} \alpha_{\textbf{k}^\prime} e^{\mathrm{i}(\textbf{k}^\prime -\textbf{k})\cdot \textbf{r}_\phi}$ where $\phi$ = a or b, corresponding to the two \ce{Cr^{3+}} sublattices ($N$ = 2 and $S$ = 3/2). The bosonic Hamiltonian is then diagonalized using wavefunction $\psi_\textbf{k} = \begin{pmatrix} a_\textbf{k} \\ b_\textbf{k} \end{pmatrix}$ to extract the magnon dispersion relations and the eigenvectors.\\

\noindent \textbf{Growth of \ce{CrI3} single crystals.} The single crystals of \ce{CrI3} were grown by the chemical vapor transport method. Chromium power (99.99\% purity) and iodine flakes (99.999\%) in a 1:3 molar ratio were put into a silicon tube with a length of 200 mm and an inner diameter of 14 mm. The tube was pumped down to 0.01 Pa and sealed under vacuum, and then placed in a two-zone horizontal tube furnace. The two growth zones were raised up slowly to 903 K and 823 K for 2 days, and were then held there for another 7 days. Shiny, black, plate-like crystals with lateral dimensions up to several millimeters can be obtained from the growth. In order to avoid degradation, the \ce{CrI3} crystals were stored in a glovebox filled with nitrogen. \\

\noindent \textbf{Fabrication of few-layer samples.} \ce{CrI3} samples were exfoliated in a nitrogen-filled glovebox and the thickness of the flakes was first estimated by the optical contrast. Using a polymer-stamping technique inside the glove box, \ce{CrI3} flakes were sandwiched between two few-layer hBN flakes to avoid surface reaction with oxygen and moisture in the ambient environment. The encapsulated \ce{CrI3} samples were then moved out of the glove box for Raman spectroscopy measurements. After Raman spectroscopy measurements, the thicknesses of the encapsulated \ce{CrI3} flakes were determined by the atomic force microscopy (AFM) measurements. \\

\noindent \textbf{Raman spectroscopy.} Raman spectroscopy measurements were carried out using both a 633 nm and a 532 nm excitation laser with a beam spot size of $\sim\SI{3}{\micro\meter}$. The laser power was kept at \SI{80}{\micro\watt}, corresponding to a similar fluence used in literature (\SI{10}{\micro\watt} over a $\sim\SI{1}{\micro\meter}$ diameter area), to minimize the local heating effect. Backscattering geometry was used. The scattered light was dispersed by a Horiba Labram HR Raman spectrometer and detected by a thermoelectric cooled CCD camera. Selection rule channels XX and XY denote the parallel and cross polarizations of incident and scattered light at $\varphi=\SI{0}{\degree}$; $\mathrm{X^{\prime}X^{\prime}}$ and $\mathrm{X^{\prime}Y^{\prime}}$ represent the parallel and cross channels at $\varphi=\SI{45}{\degree}$. A closed-cycle helium cryostat was interfaced with the micro-Raman system for the temperature dependence measurements. All thermal cycles were performed at a base pressure lower than $7\times10^{-7}$ Torr. \\


\bibliographystyle{apsrev4-1}
\nocite{apsrev41Control}
\bibliography{arXiv.bib} 

\newpage
\noindent\textbf{Acknowledgements} \\
\noindent We acknowledge useful discussions with Roberto Merlin, Lu Li, Andrew Millis and Allan McDonald. We thank Gaihua Ye for his technical assistance. R. H. acknowledges support by NSF CAREER Grant No. DMR-1760668 (Z. Y., F. D., S. S., E. W. and R. H.), and NSF MRI Grant No. DMR-1337207 (low-temperature equipment). L. Z acknowledges support by NSF CAREER Grant No. DMR-1749774 (S. L and L. Z). A. W. T acknowledges support from NSERC Discovery grant RGPIN-2017-03815 and the Korea - Canada Cooperation Program through the National Research Foundation of Korea (NRF) funded by the Ministry of Science, ICT and Future Planning (NRF-2017K1A3A1A12073407). This research was undertaken thanks in part to funding from the Canada First Research Excellence Fund. H. L. acknowledges support by the National Key R\&D Program of China (Grant No. 2016YFA0300504), the National Natural Science Foundation of China (Grants No. 11574394, No. 11774423), and the Fundamental Research Funds for the Central Universities, and the Research Funds of Renmin University of China (15XNLF06, 15XNLQ07, 18XNLG14). K. S. acknowledges support via NSF Grant No. NSF-EFMA-1741618 and the Alfred P. Sloan Foundation.\\

\noindent\textbf{Author contributions}\\
\noindent W. J. and L. Z. conceived and initiated this project; C. L., S. T. and H. L. synthesized and characterized the bulk \ce{CrI3} single crystals; H. H. K., B. Y. and A. W. T. fabricated and characterized the few-layer samples; Z. Y., P. R., F. D., S. S., E. W. and R. H. performed the Raman measurements on \ce{CrI3} thin layers, and W. J. and S. L. took the Raman spectra on bulk \ce{CrI3}; S. L. carried out the magnon dispersion calculations under the guidance of K. S. and L. Z.; W. J., R. H. and L. Z. analyzed the data; W. J., R. H. and L. Z. wrote the manuscript and all authors participated in the discussions of the results.\\

\noindent\textbf{Competing interests}\\
\noindent The authors declare no competing interests.

\newpage
\begin{figure}
\includegraphics[scale=1.2]{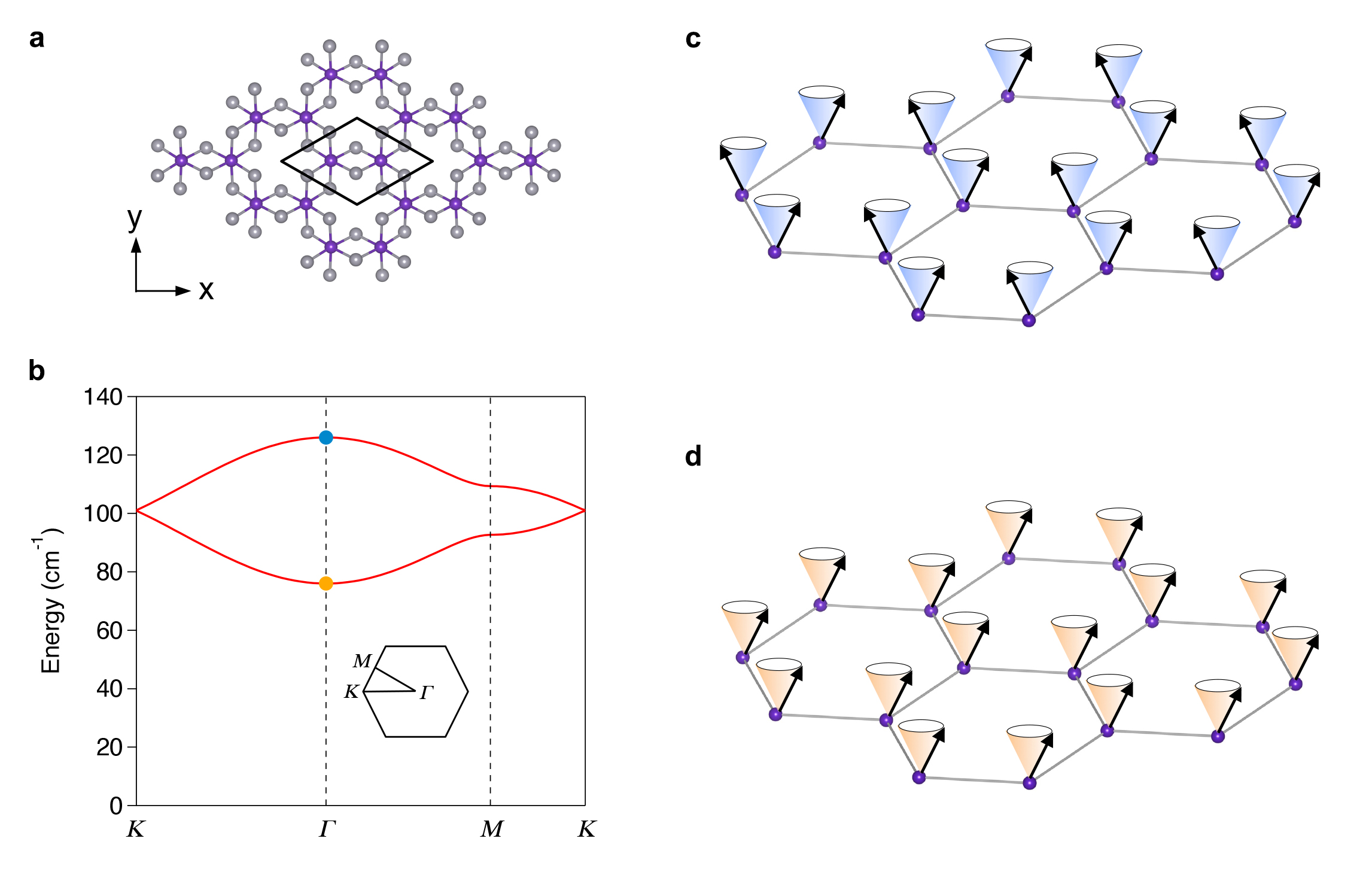}
\end{figure}
\begin{footnotesize}
\noindent \textbf{Figure 1. Magnetic excitations in the 2D Ising ferromagnet \ce{CrI3}.} \textbf{a}, Top view of the atomic structure of monolayer \ce{CrI3}, with \ce{Cr^{3+}} in purple and \ce{I^{-}} in gray. \ce{Cr^{3+}} ions form a honeycomb lattice with two \ce{Cr^{3+}} ions per unit cell (black enclosure). \textbf{b}, Calculated magnon dispersion relations in monolayer \ce{CrI3}. Inset shows the Brillouin zone. Magnon modes at the zone center ($\varGamma$ point, zero-momentum) are highlighted by yellow and blue solid circles. \textbf{c-d}, Schematic of the zone center magnon modes. The cones represent the precession trajectories of the spins (black arrows). The precession of spins on two \ce{Cr^{3+}} sublattices are out-of-phase (\textbf{c}) and in-phase (\textbf{d}), which corresponds to the high- and low-energy modes, respectively, in \textbf{b}. 
\end{footnotesize}

\newpage
\begin{figure}
\includegraphics[scale=1.2]{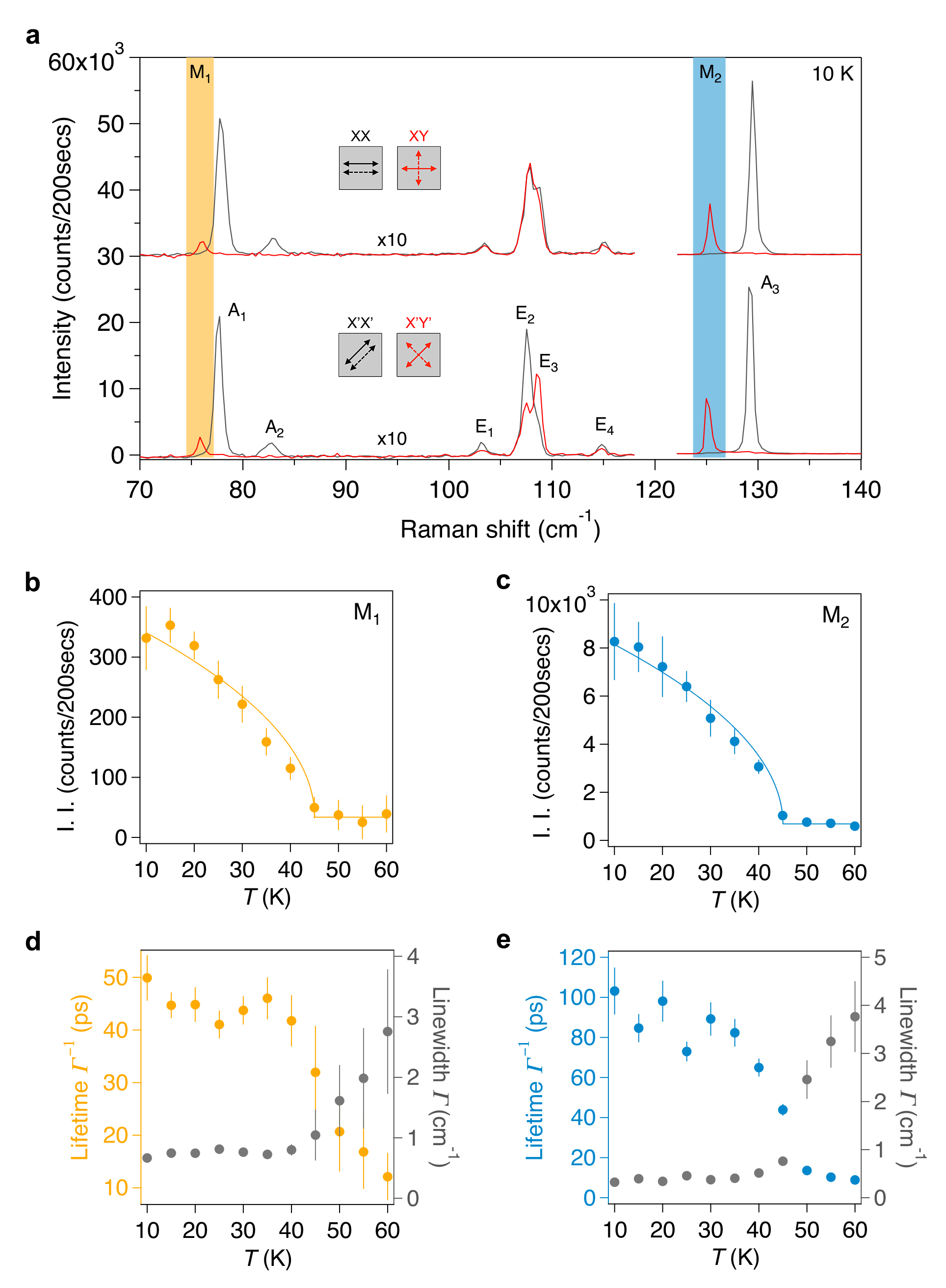}
\end{figure}
\vspace{-10pt}
\begin{footnotesize}
\noindent \textbf{Figure 2. Detection of the zero-momentum magnons in thick \ce{CrI3}.} \textbf{a}, Raman spectra of a thick \ce{CrI3} flake (13 layers) at low temperature (10 K) in the parallel and cross channels at $\varphi=\SI{0}{\degree}$ (XX and XY) and at $\varphi=\SI{45}{\degree}$ ($\mathrm{X^{\prime}X^{\prime}}$ and $\mathrm{X^{\prime}Y^{\prime}}$). Magnon modes, $\mathrm{M_1}$ and $\mathrm{M_2}$, appearing only in the cross channels (XY and $\mathrm{X^{\prime}Y^{\prime}}$), are highlighted in yellow and blue. Phonon modes are labeled as $\mathrm{A_1}$, $\mathrm{A_2}$, $\mathrm{E_1}$-$\mathrm{E_4}$, and $\mathrm{A_3}$. The spectral intensities in the 70-120 $\mathrm{cm}^{-1}$ range are multiplied by a factor of 10. The spectra in the XX and XY channels are vertically offset for clarity. The spectra are acquired using a 633 nm excitation laser. \textbf{b-c}, Temperature dependence of I. I. of the $\mathrm{M_1}$ and $\mathrm{M_2}$ magnon modes, respectively. Solid curves are fits to $I_0+I\sqrt{T_\mathrm{C}-T}$. \textbf{d-e}, Temperature dependence of the lifetime ($\varGamma^{-1}$, left axis) and the linewidth ($\varGamma$, right axis) of the $\mathrm{M_1}$ and $\mathrm{M_2}$ magnon modes, respectively. Error bars in \textbf{b-e} represent the two standard errors of fitting parameters in the Lorentzian fits to individual temperature dependent Raman spectra.
\end{footnotesize}

\newpage
\begin{figure}
\includegraphics[scale=1.2]{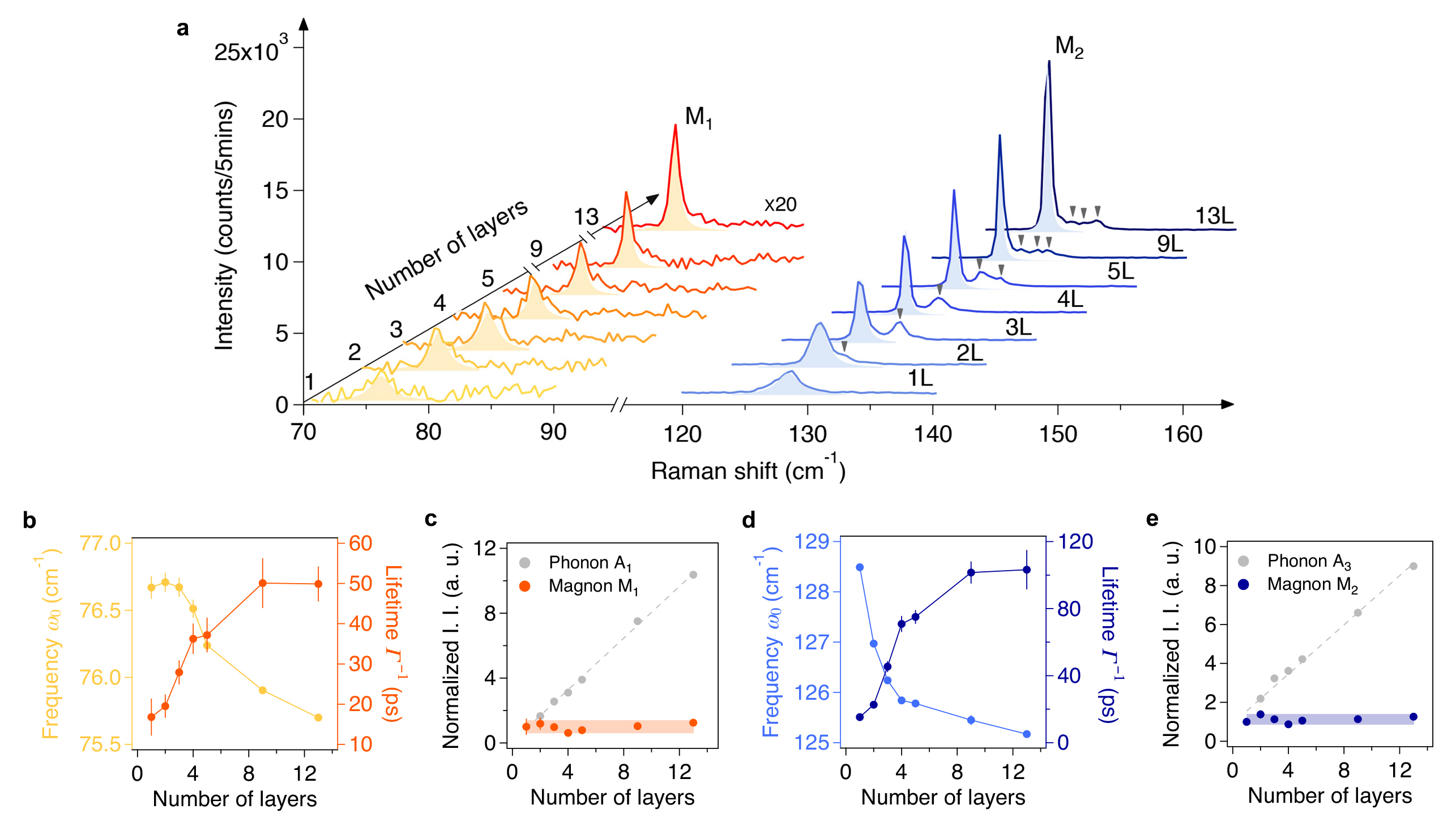}
\end{figure}
\begin{footnotesize}
\noindent\textbf{Figure 3. Layer number dependence of the zero-momentum magnon characteristics.} \textbf{a}, Raman spectra of the $\mathrm{M_1}$ (intensity multiplied by 20) and $\mathrm{M_2}$ magnons for 1-5L, 9L and 13L samples. Light yellow and blue shaded areas are the Lorentzian fits for the $\mathrm{M_1}$ and $\mathrm{M_2}$ magnon modes respectively. The gray triangles highlight the satellite magnon peaks.  \textbf{b}, Frequency ($\omega_0$, left axis) and lifetime ($\varGamma^{-1}$, right axis) of the $\mathrm{M_1}$ magnon as a function of layer number. \textbf{c}, I. I. of the $\mathrm{M_1}$ magnon as a function of layer number with I. I. of the $\mathrm{A_1}$ phonon shown in gray for comparison. \textbf{d-e}, Plots for $\mathrm{M_2}$ that are similar to \textbf{b-c}, with I. I. of the $\mathrm{A_3}$ phonon plotted in gray in \textbf{e} for comparison. Error bars in \textbf{b-e} represent the two standard errors of fitting parameters in the Lorentzian fits to individual Raman spectra at different thicknesses.
\end{footnotesize}

\newpage
\begin{figure}
\includegraphics[scale=1.3]{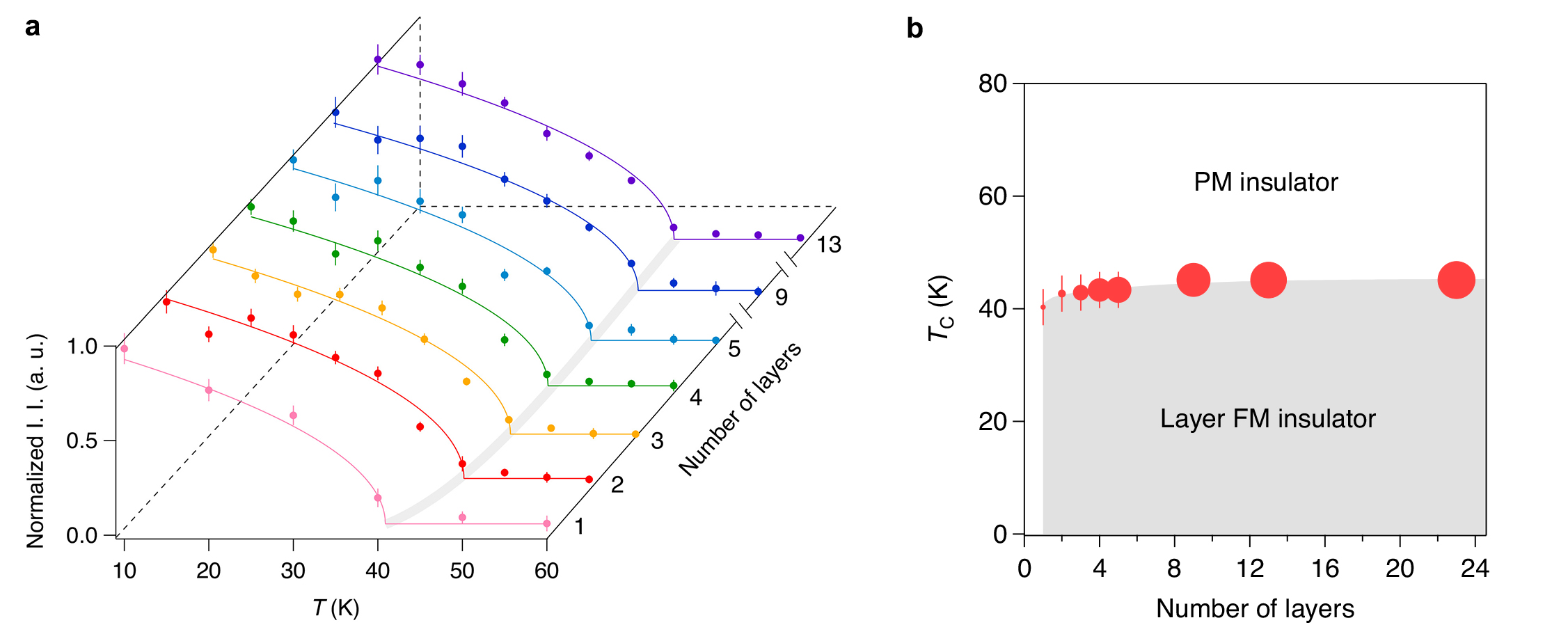}
\end{figure}
\begin{footnotesize}
\noindent\textbf{Figure 4. Temperature versus layer number phase diagram of the 2D layer ferromagnetic \ce{CrI3}.} \textbf{a}, Temperature dependence of I. I. of the $\mathrm{M_2}$ magnon normalized to the value at 10 K as a function of layer number. The coloured solid curves are fits to $I_0+I\sqrt{T_\mathrm{C}-T}$. The gray curve is the guide to the eye of the evolution of $T_\mathrm{C}$. Error bars correspond to the two standard errors of normalized I. I. in the Lorentzian fits at individual temperatures for different thicknesses of 2D \ce{CrI3}. \textbf{b}, Temperature versus layer number phase diagram (PM for paramagnetism and FM for ferromagnetism). The size of the data points represents the $\mathrm{M_2}$ magnon lifetime. Error bars correspond to the two standard errors of $T_\mathrm{C}$ in the fits in \textbf{a}. 
\end{footnotesize}

\newpage

\begin{center}
\textbf{Supplementary Information}\\

\vspace{12pt}
\textbf{Raman Fingerprint of Two Terahertz Spin Wave Branches in A Two-Dimensional Honeycomb Ising Ferromagnet}\\

Wencan Jin$,^{1,\ast}$ Hyun Ho Kim$,^{2,\ast}$ Zhipeng Ye$,^3$ Siwen Li$,^1$ Pouyan Rezaie$,^3$ Fabian Diaz$,^3$ Saad Siddiq$,^3$ Eric Wauer$,^3$ Bowen Yang$,^2$ Chenghe Li$,^4$ Shangjie Tian$,^4$ Kai Sun$,^1$ Hechang Lei$,^4$ Adam W. Tsen$,^2$ Liuyan Zhao$,^{1, \dagger}$ and Rui He$^{3, \ddagger}$\\

\textit{$^1$Department of Physics, University of Michigan, 450 Church Street,\\ Ann Arbor, Michigan 48109, USA}\\

\textit{$^2$Institute for Quantum Computing, Department of Chemistry, \\and Department of Physics and Astronomy, University of Waterloo,\\ Waterloo, 200 University Ave W, Ontario N2L 3G1, Canada}

\textit{$^3$Department of Electrical and Computer Engineering, 910 Boston Avenue, \\Texas Tech University, Lubbock, Texas 79409, USA}

\textit{$^4$Department of Physics and Beijing Key Laboratory of \\Opto-electronic Functional Materials \& Micro-nano Devices, \\Renmin University of China, Beijing 100872 China}
\end{center}

\vspace{20pt}

\noindent \textbf{Table of Contents}\\
\noindent{S1. Thickness characterization of \ce{CrI3} thin layers}\\
\noindent{S2. Comparison between on and off-resonance Raman spectra}\\
\noindent{S3. Symmetry analysis on Raman active phonons}\\
\noindent{S4. Magnetization and Raman data from bulk \ce{CrI3}}\\
\noindent{S5. Satellite phonon modes arising from the finite thickness effect}\\
\noindent{S6. Temperature dependence of $\mathrm{M}_1$ as a function of thickness}\\

\vspace{1 in}

\noindent{$\ast$ These authors contribute equally to this work.}\\
\noindent{$\dagger$ lyzhao@umich.edu}\\
\noindent{$\ddagger$ rui.he@ttu.edu}


\newpage

\noindent\textbf{S1. Thickness characterization of \ce{CrI3} thin layers}\\

\noindent Figure S1 shows the characterization of the thickness of a representative \ce{CrI3} flake on which the Raman spectra of 2-5L \ce{CrI3} shown in the main text were acquired. As stated in the Methods section, the thickness of the sample was first estimated by the optical contrast (Figure S1a and b), and then fully encapsulated by hBN flakes (Figure S1c). To prevent potential damage, the height profiles for the various thicknesses were measured by ambient atomic force microscopy (AFM) after Raman spectroscopy measurements. As a monolayer \ce{CrI3} film between two hBN flakes was determined to be 0.7 nm, this flake contains regions with four different thicknesses, including 1.4 nm (2L), 2.1 nm (3L), 2.8 nm (4L) and 3.6 nm (5L) as shown in Figure S1d-g.

\begin{figure}[!h]
\includegraphics[scale=0.6]{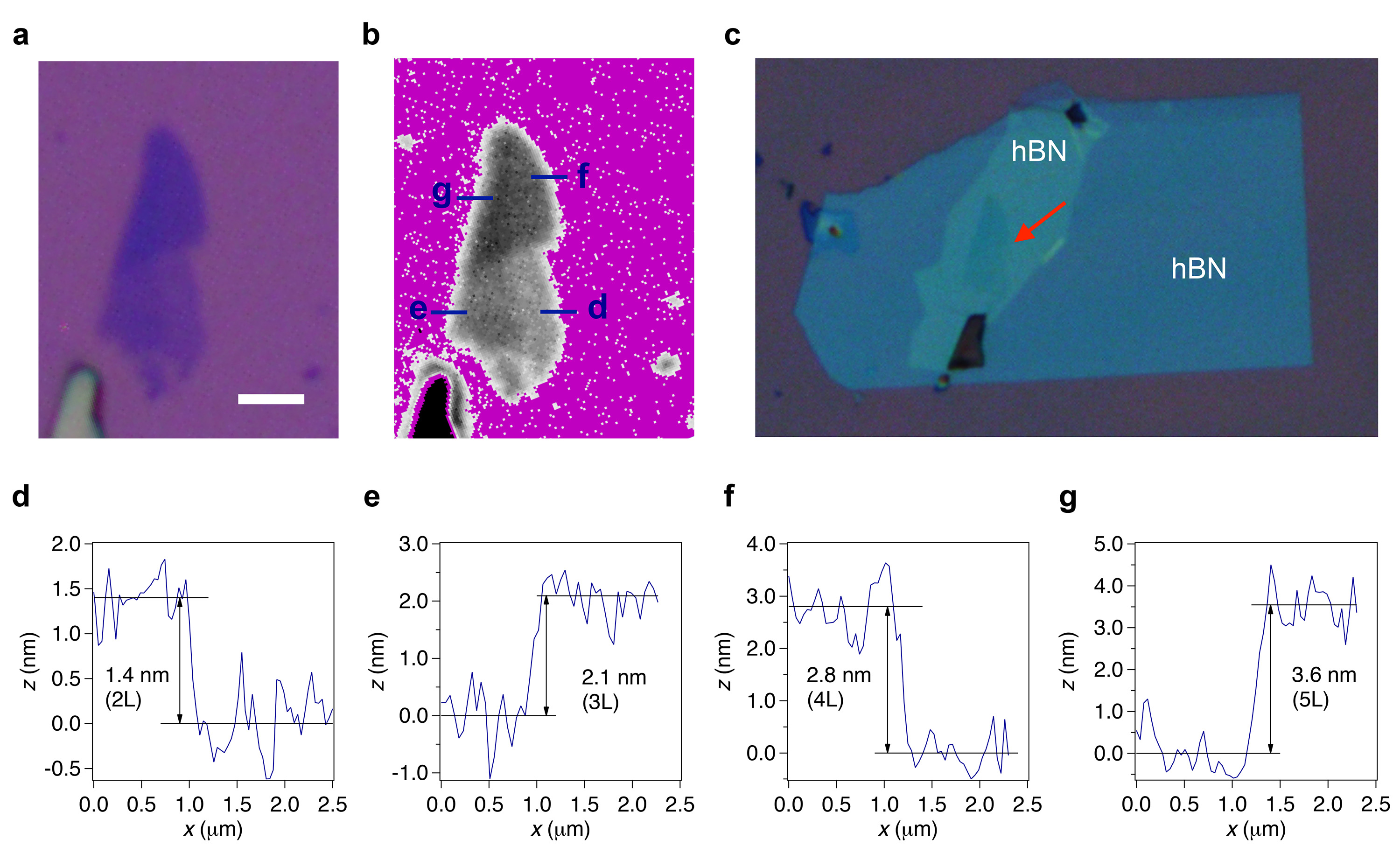}
\end{figure}
\vspace{-5pt}
\begin{footnotesize}
\noindent \textbf{Figure S1. Thickness characterization of \ce{CrI3} flakes.} \textbf{a}, Optical microscope image of a representative \ce{CrI3} flake. Scale bar is \SI{5}{\micro\meter}. \textbf{b}, Pseudocolor map of the same flake for a better visualization of the optical contrast. \textbf{c}, Optical microscope image of the same flake (marked by the red arrow) sandwiched between two few-layer hBN flakes. \textbf{d-g}, AFM line cuts across the flake edges marked in \textbf{b} with the corresponding thicknesses and layer numbers labeled.\\
\end{footnotesize}

\newpage
\noindent\textbf{S2. Comparison between on and off-resonance Raman spectra}\\

\noindent Raman spectra from the 13L \ce{CrI3}, acquired using both 633 nm (on-resonance) and 532 nm (off-resonance) excitation lasers under similar measurement conditions, are shown in Figure S2. Comparing to the off-resonance Raman spectra, the resonant Raman intensities of the $\mathrm{M}_2$ magnon mode and the $\mathrm{A}_3$ phonon mode are significantly enhanced by a factor of $\sim$20 while the rest modes are enhanced by $\sim$5. Note that the $\mathrm{M}_1$ and $\mathrm{M}_2$ magnon modes appear in the off-resonance spectra with the identical selection rules as in the resonant spectra, which rules out the possibility that the $\mathrm{M}_1$ and $\mathrm{M}_2$ magnon modes are resonance-induced symmetry forbidden phonon modes. A new phonon mode (labeled as $\mathrm{E}_4$. in Fig. 2 of the main text) appears at $\sim$115 $\mathrm{cm}^{-1}$ in the resonant spectra. A detailed analysis of $\mathrm{E}_4$ mode is beyond the scope of this paper and will be discussed in a separate work.  

\begin{figure}[!h]
\includegraphics[scale=1.3]{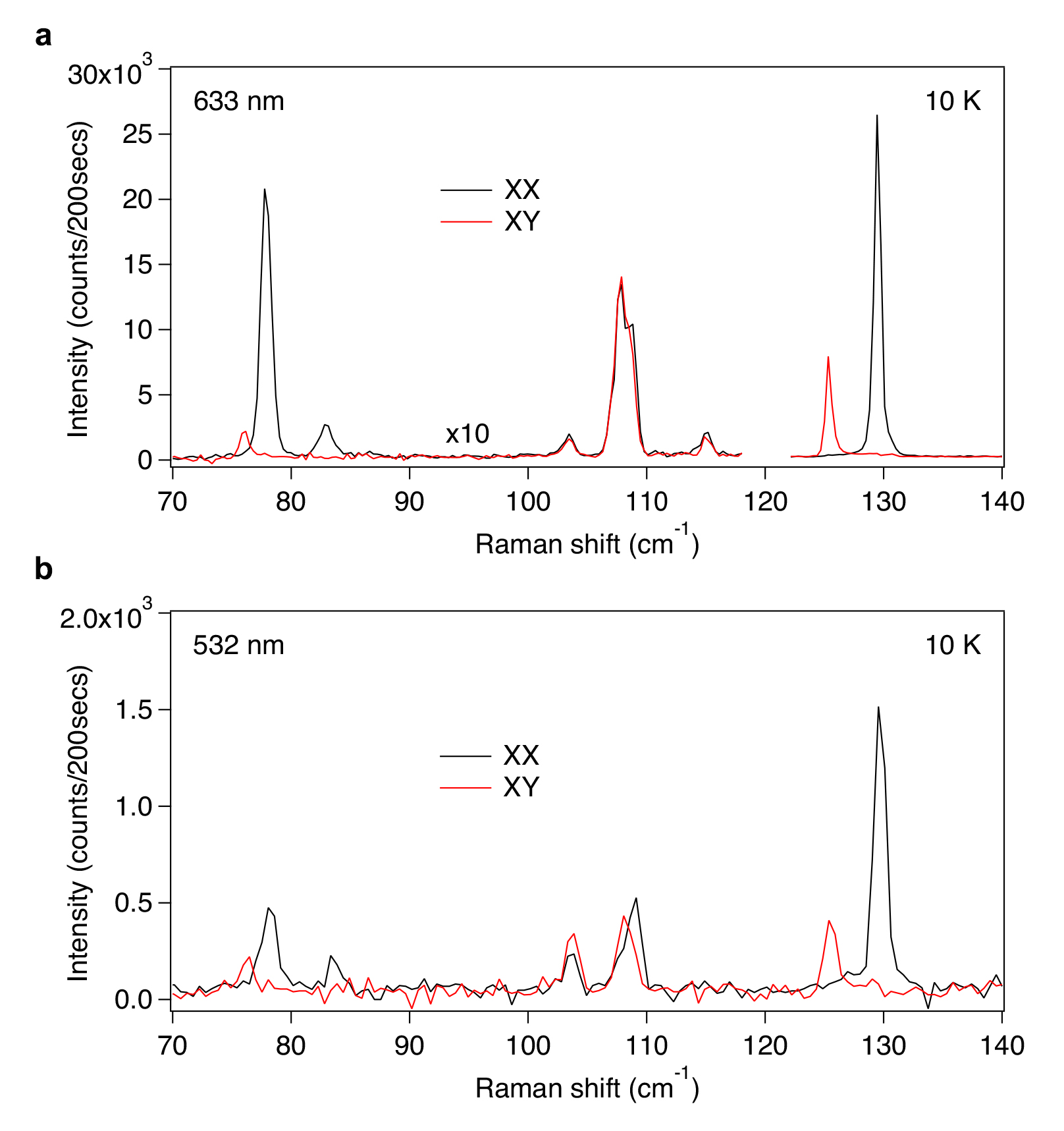}
\end{figure}
\vspace{-5pt}
\begin{footnotesize}
\noindent\textbf{Figure S2. Comparison between on and off-resonance Raman spectra.} Raman spectra from a 13L \ce{CrI3} flake in the XX and XY channels acquired at low temperature (10 K) using \textbf{a}, a 633 nm and \textbf{b}, a 532 nm excitation laser. \\
\end{footnotesize}

\newpage
\noindent\textbf{S3. Symmetry analysis on Raman active phonons}\\

\noindent At low temperature (\textit{T} $<$ 240 K), \ce{CrI3} crystal has a Rhombohedral structure (point group $C_\mathrm{3i}$ and space group $R\bar{3}$). A factor group analysis reveals that 21 optical modes are expected with irreducible representation $\varGamma_\mathrm{optical}$  = 4$A_\mathrm{g}$ + 4$E_\mathrm{g}$ + 3$A_\mathrm{u}$ + 3$E_\mathrm{u}$. Among them, $A_\mathrm{g}$ and $E_\mathrm{g}$ modes are Raman active, whose Raman tensors are given by
\begin{center}
$\chi(A_\mathrm{g})=\begin{pmatrix} a & 0 & 0 \\ 0 & a & 0 \\  0 & 0 & b \end{pmatrix}$, \hspace{1in} $\chi(E_\mathrm{g})=\begin{pmatrix} c & d & e \\ d & -c & f \\  e & f & 0 \end{pmatrix}$
\end{center}

\noindent The Raman intensity is $I \propto |\langle E_{i} | \chi | E_{s}\rangle|^2$, where $E_\mathrm{i}$ and $E_\mathrm{s}$ are the electric field of the incident and scattered light, respectively. In the backscattering geometry, the angular dependence of the Raman intensity in the parallel and cross channels is shown in Figure S3. Here, $A_\mathrm{g}$ modes only appear in the parallel channel with isotropic Raman intensities, while $E_\mathrm{g}$ modes can be observed in both parallel and cross channels with anisotropic Raman intensities. Therefore, by comparing the selection rules of the measured Raman spectra with the symmetry analysis above, we can unambiguously assign $\mathrm{A}_\mathrm{1}$, $\mathrm{A}_\mathrm{2}$ and $\mathrm{A}_\mathrm{3}$ mode to be $A_\mathrm{g}$ modes, and $\mathrm{E}_\mathrm{1}$-$\mathrm{E}_\mathrm{4}$ modes to $E_\mathrm{g}$ modes.

\begin{figure}[!h]
\includegraphics[scale=0.55]{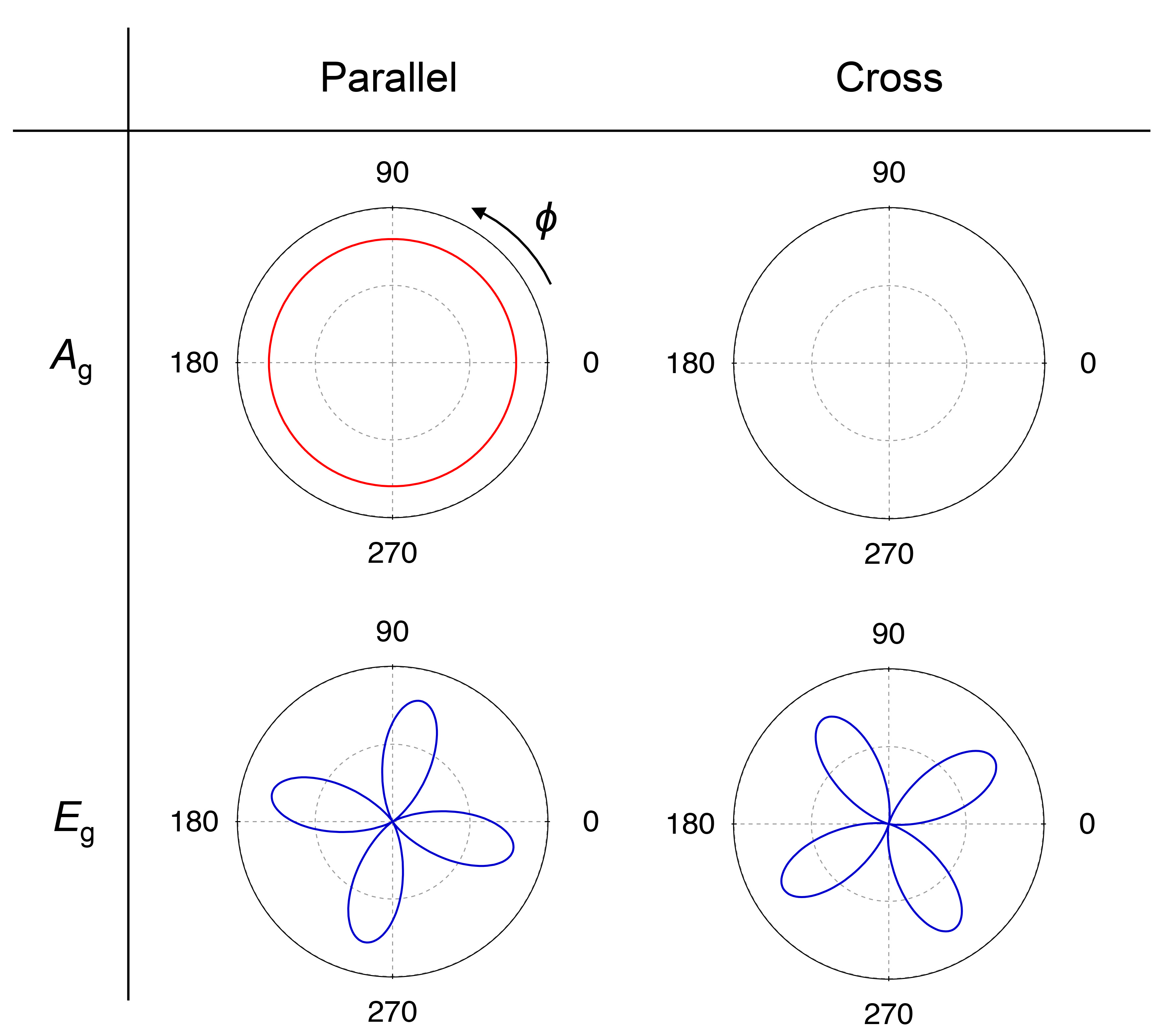}
\end{figure}
\vspace{-5pt}
\begin{footnotesize}
\noindent\textbf{Figure S3. Simulated angular dependence of the Raman intensities.} Polar plots of angular dependent Raman intensities for $A_\mathrm{g}$ and $E_\mathrm{g}$ modes in the parallel and cross channels. The polarization angle with respect to the horizontal axis is denoted as $\phi$. \\
\end{footnotesize}


\noindent\textbf{S4. Magnetization and Raman data from bulk \ce{CrI3}}\\

\noindent Out-of-plane and in-plane magnetization measurements on \ce{CrI3} single crystals were performed with the applied magnetic fields of 0.1 T, 0.5 T and 5 T. The temperature dependence of the magnetization clearly exhibits the ferromagnetic nature of \ce{CrI3}, and the Curie temperature of 60 K is determined from the lowest field (0.1 T, Figure S4a) data. The magnetization becomes nearly isotropic at a magnetic field as high as 5 T (Figure S4c), while significant magnetic anisotropy is observed below Curie temperature at lower fields (0.1 and 0.5 T, Figure S4a-b).  

\begin{figure}[!h]
\includegraphics[scale=0.7]{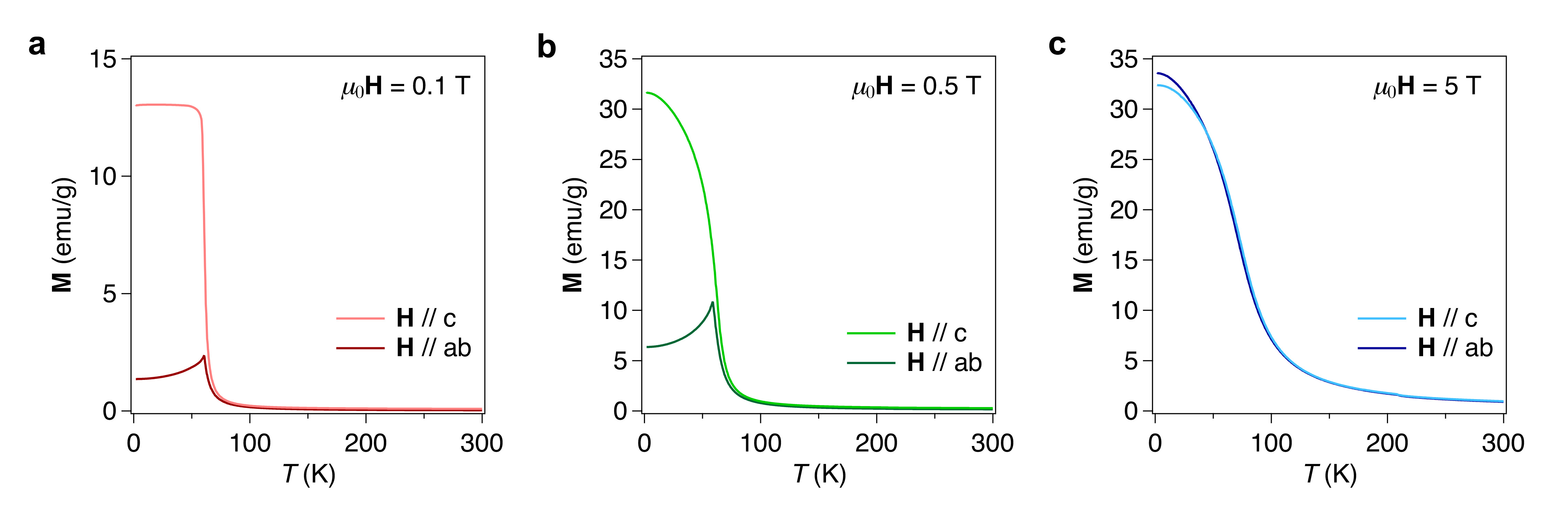}
\end{figure}
\vspace{-5pt}
\begin{footnotesize}
\noindent\textbf{Figure S4. Magnetization data from bulk \ce{CrI3} crystals.} Out-of-plane (\textbf{H} $//$ c) and in-plane (\textbf{H} // ab) magnetization as a function of temperature measured with various applied magnetic fields ($\mu_0$\textbf{H}) \textbf{a}, 0.1 T, \textbf{b}, 0.5 T and \textbf{c}, 5 T. \\
\end{footnotesize}

\noindent We further performed Raman spectroscopy measurements on a freshly cleaved \ce{CrI3} bulk and a 75L thick \ce{CrI3} flake that is prepared the same way as other flakes. The result is shown in Figure S5 and S6, and summarized as follows. Both bulk and 75L \ce{CrI3} have very similar results as 13L \ce{CrI3}, including similar magnon frequency and linewidth, and similar temperature dependence and magnetic onset temperature. This indicate that the 13L \ce{CrI3} is thick enough to represent the bulk-like magnetic properties, and therefore we choose to use 13L \ce{CrI3} data in the main text for a consistent comparison with other \ce{CrI3} thin layers prepared in the same way.

\newpage

\begin{figure}[!h]
\includegraphics[scale=0.7]{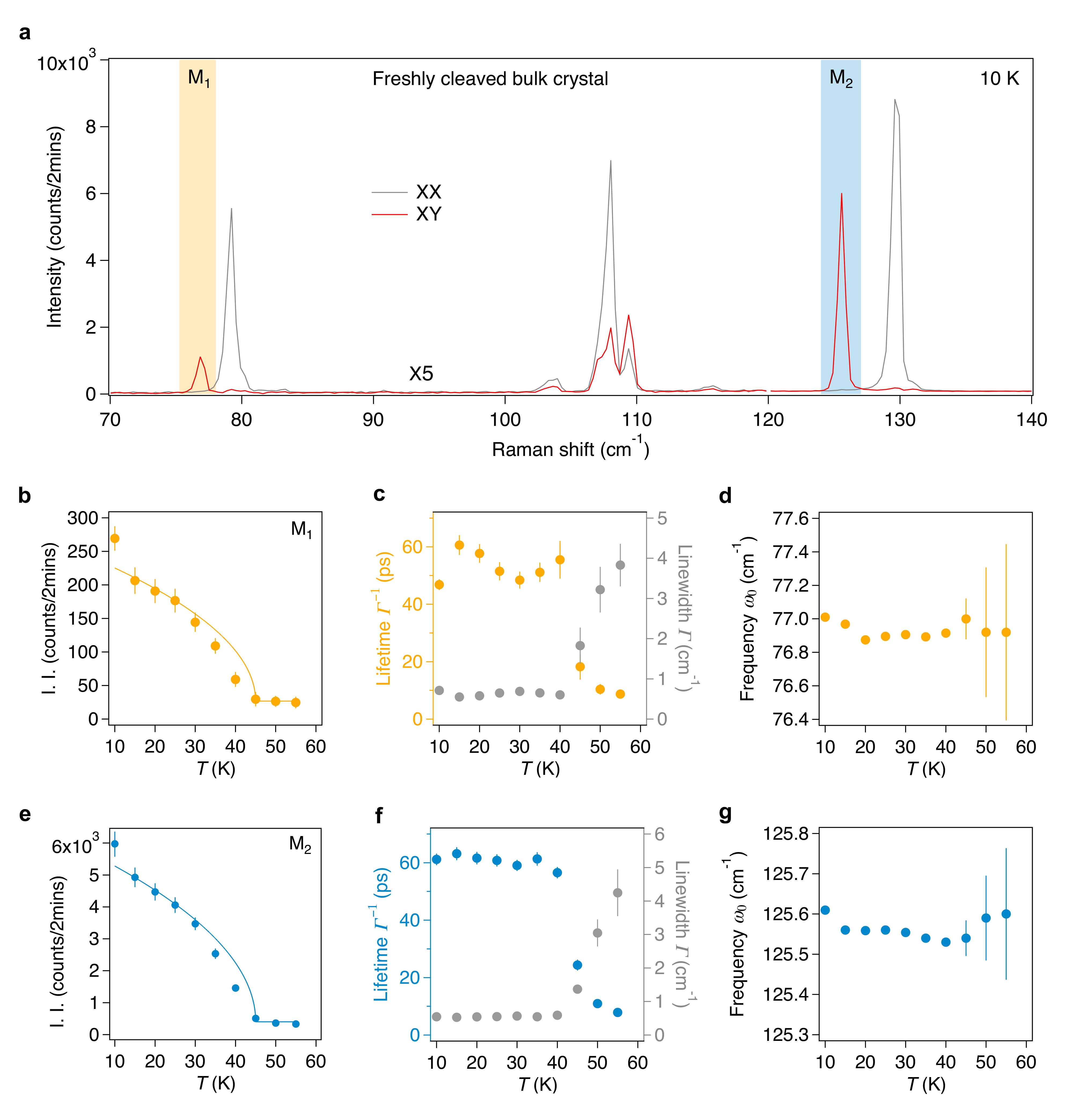}
\end{figure}
\vspace{-5pt}
\begin{footnotesize}
\noindent\textbf{Figure S5. Raman data taken on freshly cleaved bulk \ce{CrI3}.} \textbf{a}, Raman spectra of a freshly cleaved bulk \ce{CrI3} crystal at low temperature (10 K) in the parallel and cross channels at $\phi$ = \SI{0}{\degree} (XX and XY). Magnon modes, $\mathrm{M}_1$ and $\mathrm{M}_2$, appearing only in the cross channels (XY) are highlighted in yellow and blue. The spectral intensities in the 70-120 $\mathrm{cm}^{-1}$ range are multiplied by a factor of 5. The spectra are acquired using a 633 nm excitation laser. \textbf{b-d}, Temperature dependence of I. I., lifetime ($\varGamma^{-1}$, left axis) and the linewidth ($\varGamma$, right axis), and frequency of the $\mathrm{M}_1$ magnon mode, respectively. Solid curve in \textbf{b} is fit to $I_0+I\sqrt{T_\mathrm{C}-T}$. \textbf{e-g}, Same plots for $\mathrm{M}_2$ as in \textbf{b-d}. Error bars represent the two standard errors of fitting parameters in the Lorentzian fits to individual Raman spectra at different temperatures. \\
\end{footnotesize}

\newpage

\begin{figure}[!h]
\includegraphics[scale=0.7]{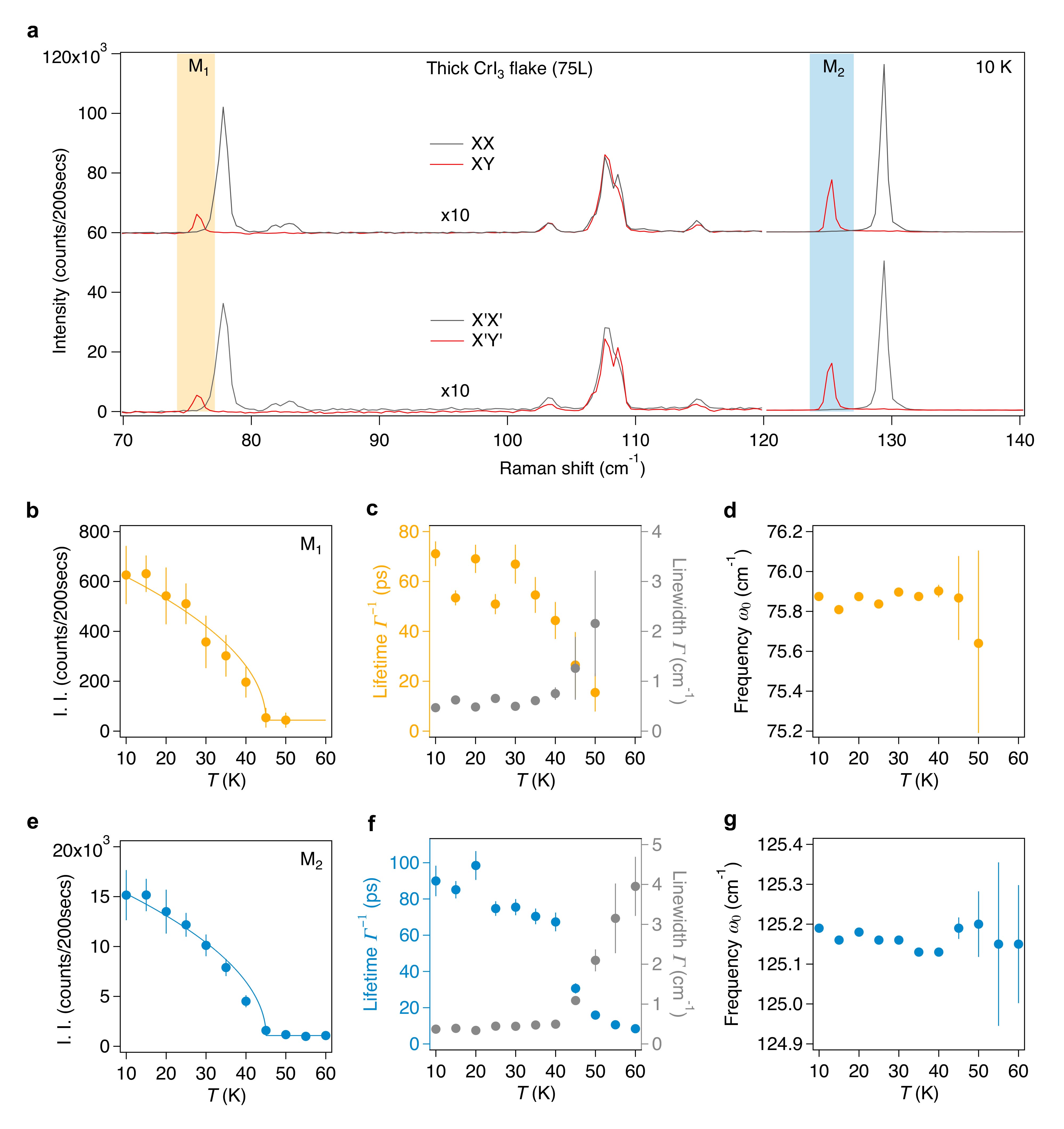}
\end{figure}
\vspace{-5pt}
\begin{footnotesize}
\noindent\textbf{Figure S6. Raman data taken on a 75L-thick \ce{CrI3} flake.} \textbf{a}, Raman spectra of a thick \ce{CrI3} flake (75L) at low temperature (10 K) in the parallel and cross channels at $\phi$ = \SI{0}{\degree} (XX and XY) and at $\phi$ = \SI{45}{\degree} ({$\mathrm{X}^{\prime}\mathrm{X}^{\prime}$} and {$\mathrm{X}^{\prime}\mathrm{Y}^{\prime}$}). Magnon modes, $\mathrm{M}_1$ and $\mathrm{M}_2$, appearing only in the cross channels (XY and {$\mathrm{X}^{\prime}\mathrm{Y}^{\prime}$}) are highlighted in yellow and blue. The spectral intensities in the 70-120 $\mathrm{cm}^{-1}$ range are multiplied by a factor of 10. The spectra are acquired using a 633 nm excitation laser. \textbf{b-d}, Temperature dependence of I. I., lifetime ($\varGamma^{-1}$, left axis) and the linewidth ($\varGamma$, right axis), and frequency of the $\mathrm{M}_1$ magnon mode, respectively. Solid curve in \textbf{b} is fit to $I_0+I\sqrt{T_\mathrm{C}-T}$. \textbf{e-g}, Same plots for $\mathrm{M}_2$ as in \textbf{b-d}. Error bars represent the two standard errors of fitting parameters in the Lorentzian fits to individual Raman spectra at different temperatures. \\
\end{footnotesize}


\newpage

\noindent\textbf{S5. Satellite phonon modes arising from the finite thickness effect}\\

\noindent Figure S7 shows the $\mathrm{A}_3$ phonon mode from 1-5L, 9L and 13L samples acquired in the parallel channel at 10 K. Satellite peaks appear at the lower frequency side of the main peak. Note that these satellite phonon peaks are at different frequencies from that of the $\mathrm{M}_2$ magnon, which rules out the possibility of the polarization leakage from the cross channel. Similar to the satellite peaks of the $\mathrm{M}_2$ magnon, the satellite peaks here come from the finite thickness effect, which are less prominent in thick sample (13L) and absent in the monolayer.

\begin{figure}[!h]
\includegraphics[scale=0.5]{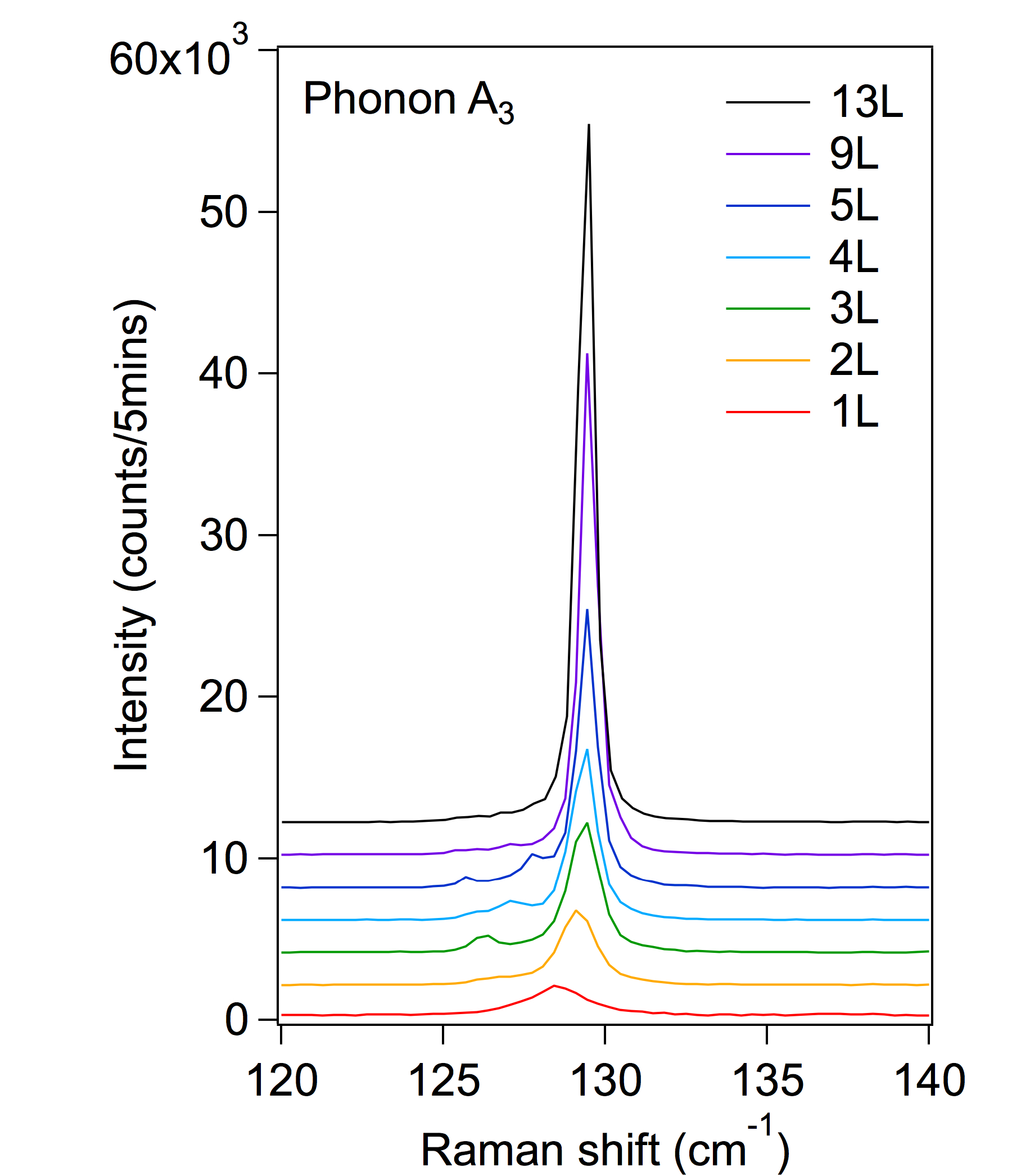}
\end{figure}
\vspace{-5pt}
\begin{footnotesize}
\noindent\textbf{Figure S7. Satellite phonon modes in \ce{CrI3} flakes with varying thicknesses. } The $\mathrm{A}_3$ phonon mode for 1-5L, 9L and 13L samples acquired in the parallel channel at 10 K. Spectra are vertically offset for clarity.  \\
\end{footnotesize}


\newpage

\noindent\textbf{S6. Temperature dependence of $\mathrm{\textbf{M}}_1$ as a function of thickness}\\

\noindent Figure S8 shows the temperature dependence of I. I. of the $\mathrm{M}_1$ magnon normalized to the value at 10 K as a function of layer number. Similar as the $\mathrm{M}_2$ magnon, normalized I. I. exhibits an order-parameter-like behavior (I. I. $\propto \sqrt{T_\mathrm{C}-T}$) as temperature approaches $T_\mathrm{C}$ from below. $T_\mathrm{C}$ determined from the $\mathrm{M}_1$ magnon is slightly lower than that of the $\mathrm{M}_2$ magnon, in particular for the monolayer sample. A potential explanation could be that the $\mathrm{M}_1$ magnon is at lower energy than the $\mathrm{M}_2$ magnon and therefore is more affected by the increased thermal fluctuations in 2D.

\begin{figure}[!h]
\includegraphics[scale=0.75]{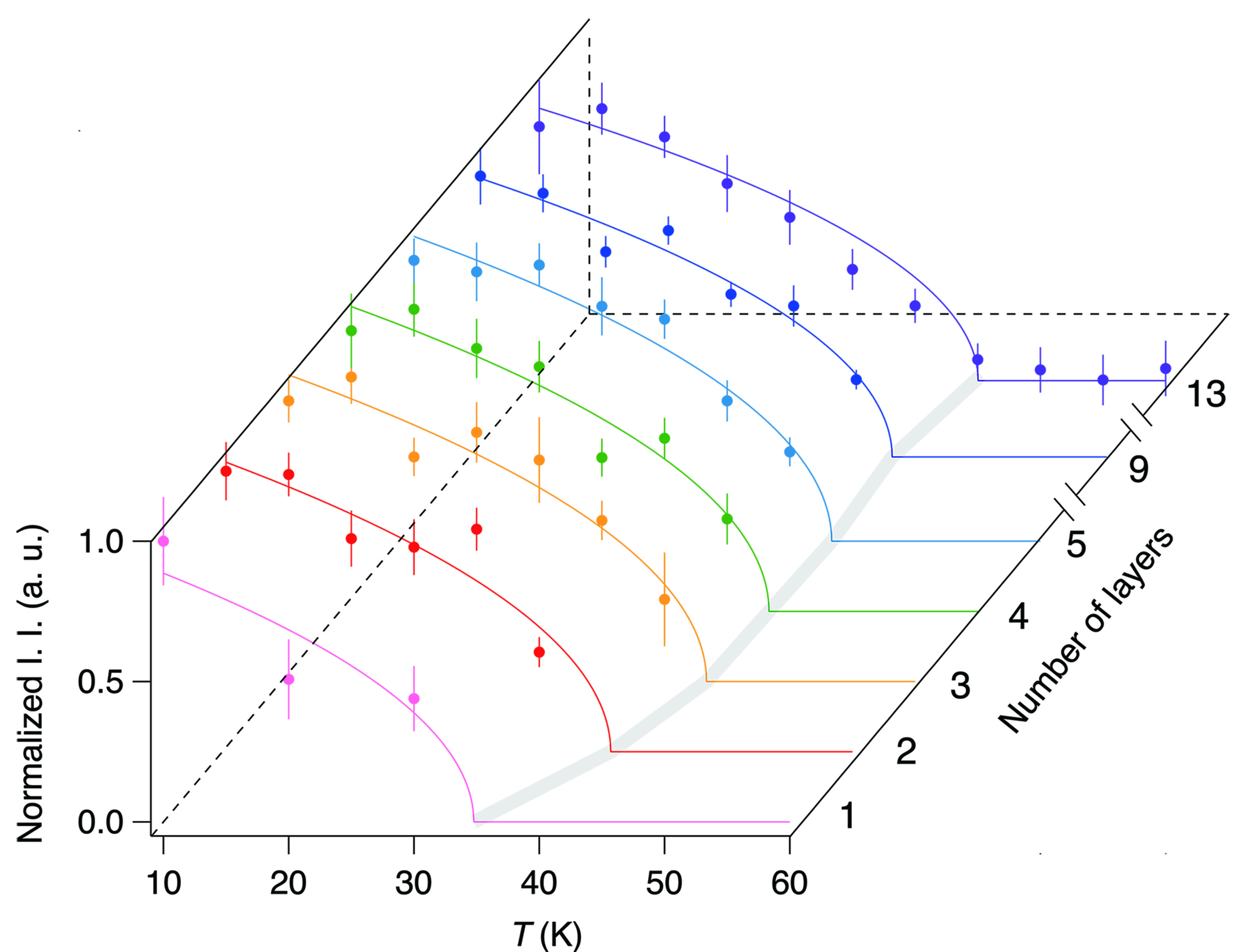}
\end{figure}
\vspace{-5pt}
\begin{footnotesize}
\noindent\textbf{Figure S8. Temperature dependence of $\mathrm{\textbf{M}}_1$ as a function of layer numbers.} Normalized I. I of the $\mathrm{M}_1$ magnon as a function of temperature for 1-5L, 9L and 13L thick samples. The gray curve is the guide to the eye of the evolution of fitted $T_\mathrm{C}$. Error bars represent the two standard errors of the normalized I. I. extracted in the Lorentzian fits of temperature dependent Raman spectra with different layer numbers.  \\
\end{footnotesize}

\end{document}

%% file: author_list.tex

\author{Wencan Jin}
\altaffiliation{These authors contribute equally to this work.}
\affiliation{Department of Physics, University of Michigan, 450 Church Street, Ann Arbor, Michigan 48109, USA}

\author{Hyun Ho Kim}
\altaffiliation{These authors contribute equally to this work.}
\affiliation{Institute for Quantum Computing, Department of Chemistry, and Department of Physics and Astronomy, University of Waterloo, Waterloo, 200 University Ave W, Ontario N2L 3G1, Canada}

\author{Zhipeng Ye}
\affiliation{Department of Electrical and Computer Engineering, 910 Boston Avenue, Texas Tech University, Lubbock, Texas 79409, USA}

\author{Siwen Li}
\affiliation{Department of Physics, University of Michigan, 450 Church Street, Ann Arbor, Michigan 48109, USA}

\author{Pouyan Rezaie}
\affiliation{Department of Electrical and Computer Engineering, 910 Boston Avenue, Texas Tech University, Lubbock, Texas 79409, USA}

\author{Fabian Diaz}
\affiliation{Department of Electrical and Computer Engineering, 910 Boston Avenue, Texas Tech University, Lubbock, Texas 79409, USA}

\author{Saad Siddiq}
\affiliation{Department of Electrical and Computer Engineering, 910 Boston Avenue, Texas Tech University, Lubbock, Texas 79409, USA}

\author{Eric Wauer}
\affiliation{Department of Electrical and Computer Engineering, 910 Boston Avenue, Texas Tech University, Lubbock, Texas 79409, USA}

\author{Bowen Yang}
\affiliation{Institute for Quantum Computing, Department of Chemistry, and Department of Physics and Astronomy, University of Waterloo, Waterloo, 200 University Ave W, Ontario N2L 3G1, Canada}

\author{Chenghe Li}
\affiliation{Department of Physics and Beijing Key Laboratory of Opto-electronic Functional Materials \& Micro-nano Devices, Renmin University of China, Beijing 100872 China}

\author{Shangjie Tian}
\affiliation{Department of Physics and Beijing Key Laboratory of Opto-electronic Functional Materials \& Micro-nano Devices, Renmin University of China, Beijing 100872 China}

\author{Kai Sun}
\affiliation{Department of Physics, University of Michigan, 450 Church Street, Ann Arbor, Michigan 48109, USA}

\author{Hechang Lei}
\affiliation{Department of Physics and Beijing Key Laboratory of Opto-electronic Functional Materials \& Micro-nano Devices, Renmin University of China, Beijing 100872 China}

\author{Adam W. Tsen}
\affiliation{Institute for Quantum Computing, Department of Chemistry, and Department of Physics and Astronomy, University of Waterloo, Waterloo, 200 University Ave W, Ontario N2L 3G1, Canada}

\author{Liuyan Zhao} 
\email{lyzhao@umich.edu}
\affiliation{Department of Physics, University of Michigan, 450 Church Street, Ann Arbor, Michigan 48109, USA}

\author{Rui He}
\email{rui.he@ttu.edu}
\affiliation{Department of Electrical and Computer Engineering, 910 Boston Avenue, Texas Tech University, Lubbock, Texas 79409, USA}